\documentclass[pdflatex,sn-mathphys-num]{sn-jnl}

\usepackage{graphicx}%
\usepackage{multirow}%
\usepackage{amsmath,amssymb,amsfonts}%
\usepackage{amsthm}%
\usepackage{mathrsfs}%
\usepackage[title]{appendix}%
\usepackage{xcolor}%
\usepackage{textcomp}%
\usepackage{manyfoot}%
\usepackage{booktabs}%
\usepackage{algorithm}%
\usepackage{algorithmicx}%
\usepackage{algpseudocode}%
\usepackage{listings}%
\usepackage{tikz}
\usepackage{tikz-3dplot}
\usepackage{subfigure}

\theoremstyle{thmstyleone}%
\theoremstyle{thmstyletwo}%

\theoremstyle{thmstylethree}%
\newcommand{\orcidbis}[1]{\href{https://orcid.org/#1}{\includegraphics[scale=0.15]{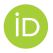}}}

\begin{document}

\title[Article Title]{The impact of seasonality over the sensitivity of Einstein Telescope and the SNR of CBC signals at the Sardinia candidate site}


\author*[1,2]{\fnm{Matteo} \sur{Di Giovanni} \orcidbis{0000-0003-4049-8336}}\email{matteo.digiovanni@sns.it}
\equalcont{These authors contributed equally to this work.}
\author[3,4]{\fnm{Davide} \sur{Rozza} \orcidbis{0000-0002-7378-6353}}\email{davide.rozza@unimib.it}
\equalcont{These authors contributed equally to this work.}
\author[5]{\fnm{Giovanni} \sur{Diaferia} \orcidbis{0000-0001-9663-0477}}\email{giovanni.diaferia@ingv.it}
\equalcont{These authors contributed equally to this work.}

\author[6]{\fnm{Andrea} \sur{Contu} \orcidbis{0000-0002-3545-2969}}

\author[7,8]{\fnm{Rosario} \sur{De Rosa} \orcidbis{0000-0002-4004-947X}}

\author[9]{\fnm{Carlo} \sur{Giunchi} \orcidbis{0000-0002-0174-324X}}

\author[10]{\fnm{Luca} \sur{Naticchioni} \orcidbis{0000-0003-2918-0730}}

\author[5,6]{\fnm{Marco} \sur{Olivieri} \orcidbis{0000-0002-7333-8809}}

\author[7,8]{\fnm{Annalisa} \sur{Allocca} \orcidbis{0000-0002-5288-1351}}

\author[7,8]{\fnm{Enrico} \sur{Calloni} \orcidbis{0000-0003-4819-3297}}

\author[6,9,11]{\fnm{Giovanni Luca} \sur{Cardello} \orcidbis{0000-0003-0521-7949}}

\author[7,8]{\fnm{Luciano} \sur{Errico} \orcidbis{0000-0003-2112-0653}}

\author[1,2]{\fnm{Giovanni} \sur{Losurdo} \orcidbis{0000-0003-0452-746X}}

\author[5]{\fnm{Irene} \sur{Molinari} \orcidbis{0000-0002-8314-1444}}

\author[8]{\fnm{Lucia} \sur{Trozzo} \orcidbis{0000-0002-8803-6715}} 

\author[6]{\fnm{Alessandro} \sur{Cardini} \orcidbis{0000-0002-6649-0298}}

\author[6,9,11]{\fnm{Domenico} \sur{D'Urso} \orcidbis{0000-0002-8215-4542}}

\affil*[1]{\orgdiv{Classe di Scienze}, \orgname{Scuola Normale Superiore}, \orgaddress{\city{Pisa}, \postcode{I-56126}, \country{Italy}}}

\affil[2]{\orgdiv{Sezione di Pisa}, \orgname{Istituto Nazionale di Fisica Nucleare}, \orgaddress{\city{Pisa}, \postcode{I-56126}, \country{Italy}}}

\affil[3]{\orgdiv{Sezione di Milano Bicocca}, \orgname{Istituto Nazionale di Fisica Nucleare}, \orgaddress{\city{Milano}, \postcode{I-20126}, \country{Italy}}}

\affil[4]{\orgdiv{Dipartimento di Fisica "Giuseppe Occhialini"}, \orgname{Università di Milano Bicocca}, \orgaddress{\city{Milano}, \postcode{I-20126}, \country{Italy}}}

\affil[5]{\orgdiv{Sezione di Bologna}, \orgname{Istituto Nazionale di Geofisica e Vulcanologia}, \orgaddress{\city{Bologna}, \postcode{I-40127}, \country{Italy}}}

\affil[6]{\orgdiv{Sezione di Cagliari}, \orgname{Istituto Nazionale di Fisica Nucleare}, \orgaddress{\city{Monserrato (Cagliari)}, \postcode{I-09042}, \country{Italy}}}

\affil[7]{\orgdiv{Dipartimento di Fisica "Ettore Pancini"}, \orgname{Universitá Federico II Napoli}, \orgaddress{\city{Napoli}, \postcode{I-80126}, \country{Italy}}}

\affil[8]{\orgdiv{Sezione di Napoli}, \orgname{Istituto Nazionale di Fisica Nucleare}, \orgaddress{\city{Napoli}, \postcode{I-80126}, \country{Italy}}}

\affil[9]{\orgdiv{Sezione di Pisa}, \orgname{Istituto Nazionale di Geofisica e Vulcanologia}, \orgaddress{\city{Pisa}, \postcode{I-56125}, \country{Italy}}}

\affil[10]{\orgdiv{Sezione di Roma 1}, \orgname{Istituto Nazionale di Fisica Nucleare}, \orgaddress{\city{Roma}, \postcode{I-00185}, \country{Italy}}}

\affil[11]{\orgdiv{Dipartimento di Chimica, Fisica, Matematica e Science Naturali}, \orgname{Università degli Studi di Sassari}, \orgaddress{\city{Sassari}, \postcode{I-07100}, \country{Italy}}}


\abstract{This work presents the first investigation of the impact of seasonal variations in seismic noise on the low-frequency performance of the Einstein Telescope (ET) at the Sardinia candidate site, focusing on their effect on compact binary coalescence signals. Using seismic data collected between 2022 and 2025 from a deep borehole, we characterize the seasonal variability of the seismic noise and identify the months of July and December as representative best- and worst-case conditions.
The measured seismic spectra are used to estimate the Newtonian noise contribution in the 2--10 Hz frequency range and to derive corresponding modifications to the ET design sensitivity. These sensitivity curves are then used in simulations of binary neutron star and intermediate-mass black hole signals in the triangular ET configuration to quantify the resulting changes in signal-to-noise ratio.
We find that median seasonal conditions produce only minor changes in the expected SNR, with deviations from the design case typically below 2$\%$. Larger effects are found for the upper tail of the December noise distribution, where the average SNR loss reaches approximately 20--25$\%$. These results indicate that seasonal seismic variability at the Sardinia candidate site has a limited impact on the low-frequency performance of ET under typical conditions.}




\maketitle
\section{Introduction}
\label{sec:Introduction}
Based on the successful exploitation \citep{gwtc1, gwtc2, gwtc3, gwtc4, gwtc42, BIGONGIARI2026} of current 2nd generation gravitational wave (GW) detectors (Advanced Virgo \citep{aVirgo}, Advanced LIGO \citep{aLIGO} and KAGRA \citep{kagra}), the scientific community is thoroughly investigating \cite{Maggiore_2020, coba, Iacovelli_2024, abac2025scienceeinsteintelescope, codazzo2026impactcoalescencesignalssearch} the scientific potential of next-generation ground based GW detectors that are expected to start observations in the early 2040s, i.e., Cosmic Explorer (CE) \cite{CE1, CE2, CE3, DiGiovanni2025_ETCE} and the Einstein Telescope (ET) \cite{et, ET2010, ET2011, ET2020, DiGiovanni2025_ETCE}, which is the subject of this work. ET was first proposed in 2010 and greatly improves the performance of current-generation detectors: the lower bound of the accessible bandwidth will be moved from the 20 Hz edge of current detectors to 2 Hz \cite{ET2020} and the sensitivity will be improved up to a factor 8 across the band covered by current detectors \cite{ET2020}. 

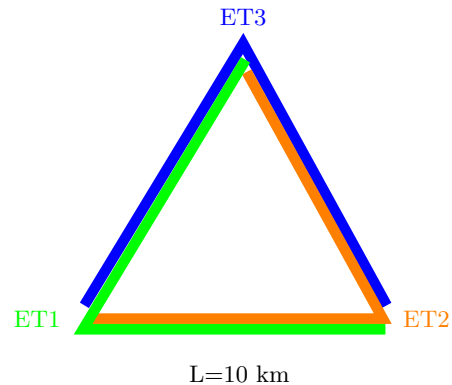
\begin{figure}[t]
\begin{tikzpicture}[scale=1pt]
    \draw [yshift=5] [line width=4] [blue] (0,0) -- (2.1,3.464) -- (4,0);
    \draw [yshift=0] [xshift=-1.5] [line width=4pt] [orange] (0.15,0) -- (4,0) -- (2.2,3.264);
    \draw [yshift=-4][xshift=-0.5] [line width=4pt] [green] (4,0) -- (0,0) -- (2.15,3.564);
    \coordinate[label={[green]left:ET1}]  (ET1) at (-0.2,0);
    \coordinate[label={[orange]right:ET2}] (ET2) at (4.1,0);
    \coordinate[label={[blue]above:ET3}] (ET3) at (2.1,3.764);
    \coordinate[label=below:{L=10 km}](c) at (2.05,-0.5);
  \end{tikzpicture}
\caption{Scheme of ET for the triangular configuration. Each detector ET1,2,3 is composed of an interferometer optimized for low-frequency detection and another optimized for high-frequency detection for a total of six.}
\label{xylophone}
\end{figure}

At present, the definitive layout of ET is still under evaluation. The most recent proposal recalls the current detector network, with two widely separated L-shaped detectors with 15 km long arms\cite{coba, Iacovelli_2024}. On the other hand, the original project foresees three pairs of nested detectors arranged in an equilateral triangle (Figure \ref{xylophone}) with the sides 10\,km long \cite{et, ET2010, ET2011, ET2020}. In both proposals, each detector is composed of an interferometer optimized for low frequencies (LF, 2\,Hz $\lesssim f \lesssim$ 40\,Hz) and another for high frequencies (HF, $f\gtrsim$ 40\,Hz). In both the 2L and the triangle configurations, ET will be hosted underground, at a currently planned depth between 200\,m and 300\,m to reduce seismic motion at the input of the suspension system of the mirrors and to reduce the impact of atmospheric disturbances \citep{hutt} and Newtonian noise (NN) \cite{Weiss:1972, beccaria1998relevance, Hughes1998SeismicGravityGradient, harms, harms2022}.

Reaching the LF target sensitivity is the most ambitious and challenging goal of ET. In fact, the extension of the bandwidth to 2\,Hz and the sharp increase in sensitivity will significantly improve the rate of detected events and give the possibility to issue early warnings for the coalescence of compact objects (CBC) several minutes, if not hours depending on the source, before the merger \cite{Branchesi_2016,Maggiore_2020,Nitz_2021,coba, Hu_2023}. The detection of CBC signals with great advance will be beneficial for multimessenger studies \cite{multim, Branchesi2018, radice, nature}, allowing for detailed observations of binary neutron star (BNS) mergers. The significant increase in sensitivity below $100 \rm\, Hz$ will also allow for more precise observations about the merger of intermediate mass black holes (IMBH) \cite{Koliopanos:2018sW, mezcua, Maggiore_2020, greene_2020}.

\begin{figure}
     \centering
     \includegraphics[width=\linewidth]{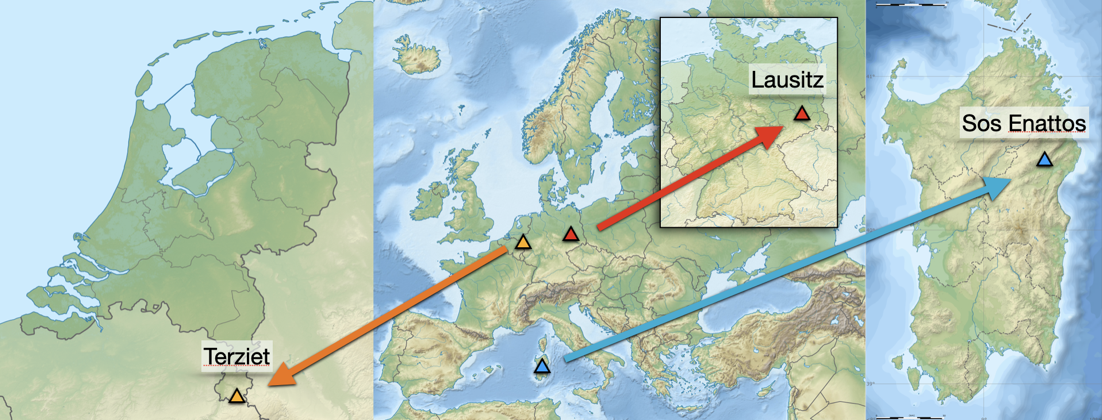}
     \caption{Map of Europe showing the locations of the official ET candidate sites. The village of Terziet in the Netherlands is usually taken as a reference for the EMR candidate site; the Sos Enattos mine is taken as a reference for the Sardinia site. In this work, we consider the Sardinia site. The maps are taken from \cite{sardiniamap, netherlandsmap, europemap} and modified according to the creative commons license 3.0. The orientation of the triangle icons is not representative of the actual proposed orientations for the ET detector at each site.}
     \label{fig:map}
 \end{figure}

As of 2026, there are three sites which are officially candidate to host the ET detector (Figure \ref{fig:map}): the  Euregio Meuse-Rhine (EMR) \cite{ET2011, ET2020}, between Belgium and the Netherlands, the area surrounding the Sos Enattos former mine in Sardinia (Italy) \cite{ET2011, ET2020, naticchionietal2014, naticchionietal2020, digiovannietal2021, digiovanni2023, saccorotti23, diaferia, Diaferia26} and the Lausitz region in Saxony (Germany). The final choice for the location of ET is expected to happen by the third quarter of 2027 and will depend on the detector configuration, which will be decided in advance before that deadline.

Previous site characterization studies for ET \cite{digiovannietal2021, digiovanni2023} showed how the potential threat from environmental noise sources of both natural \cite{higgins1950, ward1966wind, cessaro1994, withers, acernese2004properties, coward2005characterizing, virgo2006, burtin2008spectral, anthony18,   smith, dybing, o3noise, anthony} and anthropogenic \cite{acernese2004properties, virgo2006, saccorotti, piccinini, poli, o3noise, diaferia} origin can contribute as a limiting factor to the LF detector sensitivity. As a consequence, in this work we present, for the first time, an assessment of the impact of seasonal fluctuations from seismic noise levels at the Sardinia candidate site 
over the design sensitivity of ET and its scientific potential for CBC signals in the case of the triangular configuration. This is regarded as valuable information for the design of adaptive noise mitigation systems and for the engineering of the detector infrastructure. In fact, the triangular configuration poses the greatest challenges in terms of the control of environmental noise sources. Equally important, we discuss some of the properties of borehole seismic data which were missing from previous works, e.g., we rate the quietness of the Sardinia candidate site at the target depth of ET at a global scale, by ranking the recorded underground seismic noise levels against a worldwide set of seismic stations in the frequency band of interest. This is also the first study which makes use of the official proposed orientation for the ET triangle in Sardinia.

To evaluate the impact of seasonality, we follow the same approach used in Ref. \cite{DiGiovanni2025} and focus on (NN) \cite{Weiss:1972,beccaria1998relevance,Hughes1998SeismicGravityGradient,harms, harms2022} from seismic ground motion in the $[2,10]\rm\, Hz$ range, where ambient noise is more prominent \cite{digiovannietal2021, digiovanni2023, alloccaetal2021} and may affect early warnings. Beyond 10 Hz, noise contributions from the detector hardware dominate the noise budget and, therefore, are out of the scope of our work, as well as other potential noise sources from the detector's hardware, e.g., LF vertical thermal noise from the vibration isolation system. Moreover, as a consequence of the focus in the LF range, our goal is not to claim if a merger will be detected or not (merger frequencies can be beyond the considered frequency range) or if we can perfectly reconstruct a waveform, but to quantify the impact that deviations from the ET design sensitivity can have on the SNR of the signals assuming perfect knowledge of the waveforms. The details of how parameter estimation and other cosmological implications would be affected will be the subject of other works which are currently in preparation.

The paper is organized in the following way: in Section \ref{sec:background} we discuss the astrophysical background of this study; the method is presented in Section \ref{sec:Method}; Section \ref{sec:Data} talks through the data used for this study; the results are discussed in Section \ref{sec:snr}. Finally, the conclusions are presented in Section \ref{sec:conc}.


\section{Astrophysical background}
\label{sec:background}
The first source of interest for this work are IMBH \cite{mezcua,Koliopanos:2018sW, greene_2020}, which have a mass range between those formed by stellar collapse (i.e. $M < 10^2 \rm M_{\bigodot}$) and the supermassive black holes at the centers of galaxies (i.e. $M > 10^5 \rm M_{\bigodot}$). If in a binary system, their merger frequency is below 100 Hz. Before the third observing run (O3) \cite{imbho3} of aLIGO and AdVirgo, there was no clear evidence about their existence.
The strongest traces for IMBHs came from a few low-luminosity active galactic nuclei, in which the black hole masses can be estimated using the technique of reverberation mapping.\cite{Chilingarian_2018} Some ultraluminous X-Ray sources in nearby galaxies were also suspected to host IMBHs \cite{Maccarone2007} as well as the globular cluster NGC-224-G1.\cite{Baumgardt_2003} Then, due to the constant increase in the sensitivity of current GW detectors, two CBC events \cite{GW190521, GW231123} delivered the first direct evidence of the existence and formation of IMBH. A few more marginal candidates were also found, but none sufficiently significant to indicate detection of further IMBH mergers \cite{imbho3,lvkimbh}. These compact objects merge at the very low end of the sensitivity band of current detectors and last only a few milli-seconds in their accessible bandwidth, making them very hard to observe. Conversely, owing to the extension of the sensitivity down to 2 Hz, IMBH will spend up to a few minutes in ET. On top of that, the reaching of the ET LF design sensitivity goal will unlock the possibility of detailed observations of these sources as well as a significant increase in their detection rate. These observations will help to determine the formation channel of IMBH\cite{gravity, greene_2020, imbho3} and to possibly uncover part of the mystery related to the origins of dark matter \cite{pbh1}.

BNS, on the other hand, are a completely different source. They are composed of two NS, have an expected total mass between $2.2 \rm M_{\bigodot}$ and $4.4 \rm M_{\bigodot}$ \cite{salafia} and merge at frequency $\mathcal{O} (\rm kHz)$. All the relevant information carried by GW about the physics of neutron stars can be found in the HF part of the signal as well. However, LF detection plays a crucial role in the observation of BNS mergers. In fact,  detecting a BNS as soon as possible may not only significantly increase the observation rate \cite{Nitz_2021, coba, miller}, but also allow for precise sky localization \cite{Hu_2023} and sending prompt early warnings to observatories for conducting multimessengers observing campaigns \cite{Branchesi_2016, Maggiore_2020}. Moreover, BNS typically spend several hours at low frequency accumulating SNR before the chirp, since the number of cycles of the waveforms between 2 Hz and 10 Hz is $\mathcal{N}_{\rm{cyc}} ([2,10]\rm\,Hz])\simeq 7\times 10^5$, whereas the total number of cycles before merger is $\mathcal{N}_{\rm{cyc}} ([2,4096]\rm\,Hz])\simeq 7.6\times 10^5$. This means that only fully accomplishing the LF design sensitivity, especially below 10 Hz, may allow to exploit the potential of ET in multimessenger astronomy \cite{coba, Nitz_2021, DiGiovanni2025}.

\section{Methodology}
\label{sec:Method}

To estimate the impact of seasonal environmental noise \cite{digiovanni2023} over the SNR of CBC events, we use the same methodology developed for Ref. \cite{DiGiovanni2025} and 
use the design sensitivity unfolded by Ref. \cite{coba}. As already mentioned, we focus on NN from ground motion between 2 Hz and 10 Hz which we estimate analytically as \cite{harms,harms2022}
\begin{equation}\label{eq:NN}
    \tilde{h}_{\rm{NN}}(f) = \frac{4\pi}{3}G\rho_0\frac{2\sqrt{2}}{L}\frac{1}{(2\pi f)^2}\tilde{x}(f),
\end{equation}
where the frequency ($f$) dependence of the NN budget ($\tilde{h}_{\rm{NN}}$) is related to the gravitational constant ($G$), the density of the ground medium ($\rho_0$), the size of the arm-length of each interferometer ($L$) and the measured amplitude spectral density (ASD) of the horizontal component of seismic ground motion displacement ($\tilde{x}$). Equation \ref{eq:NN} holds for a spherical cavern in an infinite homogeneous medium \cite{harms2022} and the detailed derivation can be found in the Appendix. Equation \ref{eq:NN} is evaluated assuming:
\begin{itemize}
    \item underground seismic spectra representative of body waves only;
    \item one third of the body waves contribution is from compressional waves \cite{alloccaetal2021, harms2022, janssens2022, janssens2024, DiGiovanni2025};
    \item surface waves are negligible at depths of a few hundred meters;
    \item uncorrelated NN on the interferometer test masses.
\end{itemize}
In addition to that, Equation \ref{eq:NN} does not include any suppression factor, either from NN mitigation systems or from the ET infrastructural design. These aspects have not been definitively established yet, thus making any choice of the suppression factor arbitrary. Moreover, the lack of any mitigation factor in this work sets us on the conservative side. Generally speaking, the adopted curves should be understood as representative sensitivity scenarios based on the available site-dependent noise models, rather than as definitive predictions of the sensitivity that would ultimately be achieved. Indeed, the final noise budget will depend on the detailed detector design, the placement and orientation of the interferometers, further enquiry from site characterization activities.

To estimate $\tilde{x}$, we use the median of the spectral distributions of the ASD calculated on $120\rm\,s$ windows. The length of the time window is chosen according to the typical maximum duration of an IMBH and on the time segment duration proposed for CBC multiband analysis for BNS \cite{Hu_2023}. Under these assumptions, the ET noise budget that we use takes into account the NN evaluated from the seismic ground motion during a given period of the year in Sardinia.

After inferring the modified ET sensitivity curve, for consistency we use the same events considered by Ref. \cite{DiGiovanni2025}, taken from the catalog \cite{cobacatalog} used to produce the results of Ref. \cite{coba}. We then use PyCBC \cite{pycbc} to generate and inject signals in simulated ET noise. Waveforms are generated using the IMRPhenomD \cite{imrphenomd, imrphenomd2} GW approximant for IMBH and IMRPhenomPv2\_NRTidalv2 \cite{imrphenomp} for BNS, both available in the LIGO-Virgo-KAGRA Algorithm Library (LAL) \cite{lalsuite}. The total number of events considered is $\simeq$ 2e3 for IMBH and BNS, respectively. To take into account the antenna pattern of the detector, the ET triangle is simulated considering three L-shaped co-located detectors, with an arm length of $10 \rm\, km$ and an arm angle of $60^{\circ}$. Contrary to \cite{DiGiovanni2025}, which used arbitrary values, the orientation of the detectors is now determined by the coordinates of each vertex of the triangle, which have recently been established by the Sardinia local site characterization team and released here for the first time:
\begin{itemize}
    \item V1: [9.44896$^{\circ}$E, 40.472806$^{\circ}$N]
    \item V2: [9.32949$^{\circ}$E, 40.441343$^{\circ}$N]
    \item V3: [9.35350$^{\circ}$E, 40.536110$^{\circ}$N]
\end{itemize}
For each detector, noise is obtained by generating white noise colored according to the given sensitivity curve. Waveforms are then injected into the noise of each detector composing ET (labeled ET1, ET2 and ET3 respectively).

Finally, we infer the distribution of the SNR of the signals injected in simulated noise obtained using the ET design sensitivity. Signal SNR is calculated as
\begin{equation}
    SNR= \sqrt{MSNR_{\rm{ET1}}^2+MSNR_{\rm{ET2}}^2+MSNR_{\rm{ET3}}^2},
\end{equation}
where $\rm\,MSNR_{\rm{ET1,2,3}}$ are the matched filter SNR calculated in each detector. Then, using the same events, we repeat the procedure for each curve inferred from Equation \ref{eq:NN} using a given season of the year. The new SNR distributions are then compared against the design case used as a benchmark. The distributions of the SNR losses are obtained after calculating, for each event, the ratio between the new SNR and the benchmark SNR. We also assume perfect knowledge of the source, i.e., MSNR is calculated using the same waveform template injected in the data.


\section{Seismic Data}\label{sec:Data}
\begin{figure}[htbp]
     \centering
     \includegraphics[width=0.8\linewidth]{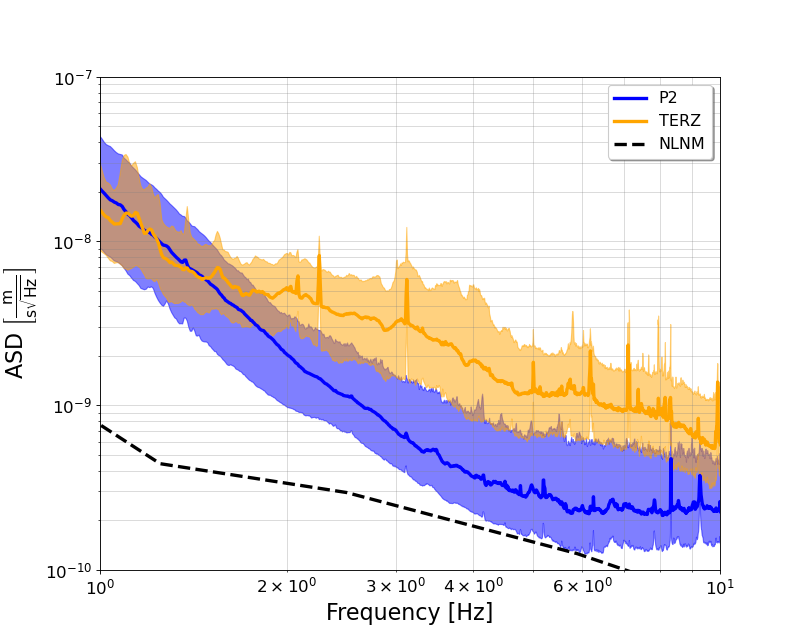}
     \caption{Spectra from borehole sensors at P2 (blue) for the period from January 2022 until July 2025. These data are considered as representative of the Sardinia candidate site. The data from a borehole sensor installed in the EMR region are also shown (orange). These data from not necessarily representative of the entire EMR candidate region. The solid lines represent the median of the distribution of 120 s long spectral segments. The colored bands show the area between the 5th and 95th percentile of the distributions. Peterson's NLNM is also shown (dashed line).}
     \label{fig:overall_spectra}
\end{figure}
\begin{figure}[htbp]
     \centering
     \includegraphics[width=\linewidth]{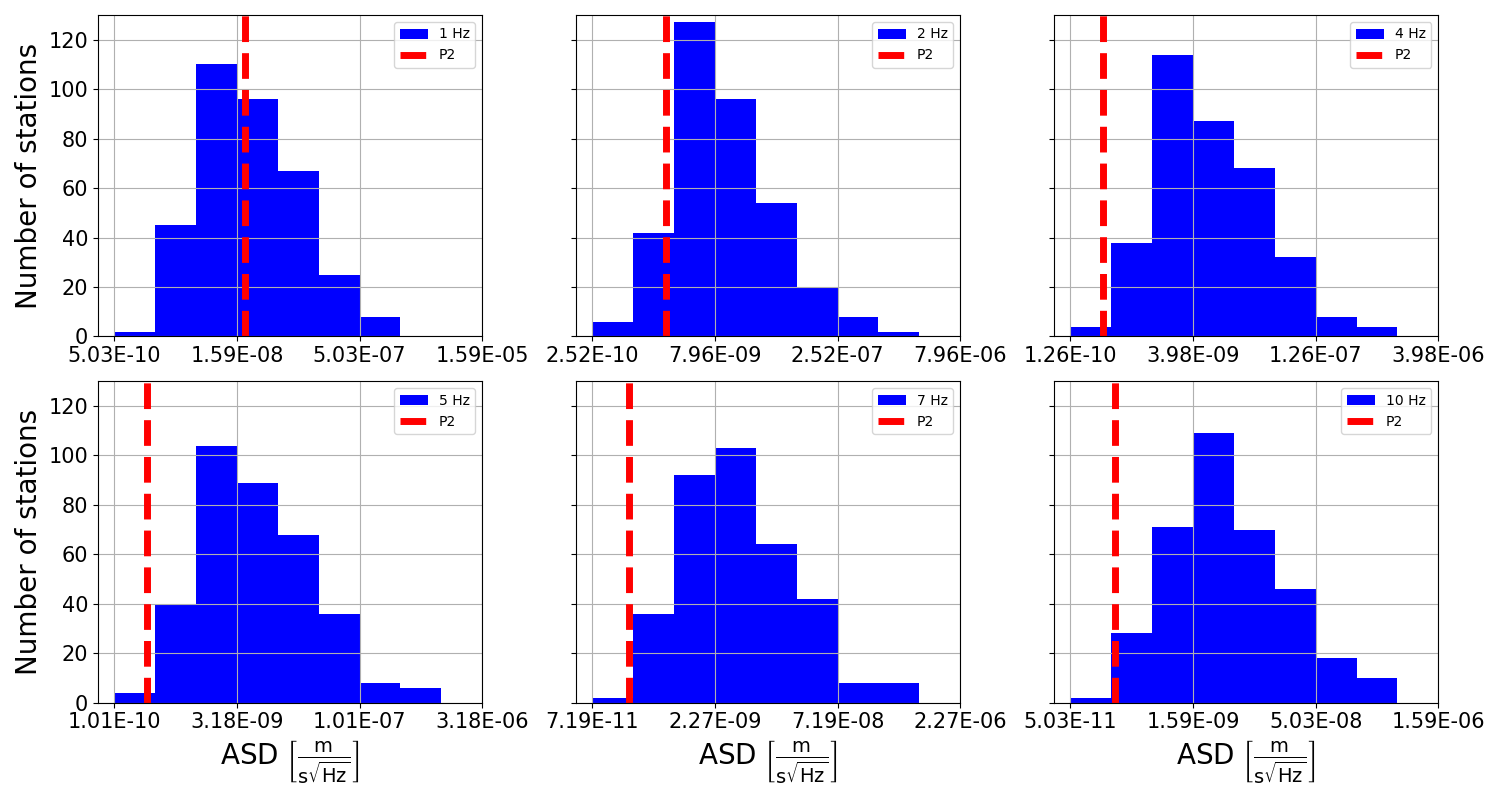}
     \caption{Evaluation of the noise level for P2 at global scale. For each frequency bin, on top of each frame, we calculate the mean ASD for each of the stations that compose the Incorporated Research Institutions for Seismology virtual network FDSN+ALL in the period from January 2022 until July 2025. In each frame, the vertical bar indicates the average level of noise as observed at P2 in the same time span.}
     \label{fig:ranking}
\end{figure}
 \begin{figure}[htbp]
     \centering
     \includegraphics[width=0.8\linewidth]{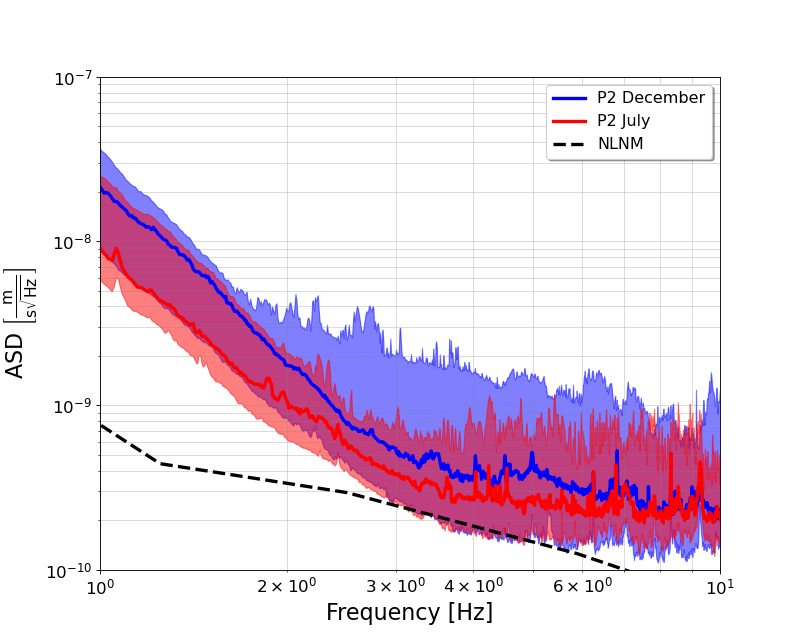}
     \caption{Comparison of the seismic spectra from P2 in July (red) and December (blue). The two represent the best and worst periods, respectively, in terms of the impact of seasonality over the seismic noise levels in Sardinia.}
     \label{fig:season_spectra}
\end{figure}
Seismic data related to the Sardinia candidate site are from a seismometer installed in borehole at $\sim -264$ at a location named P2. The depth is consistent with the expected depth of ET and the data are considered representative of the entire Sardinia candidate site. The seismometer is a Trillium 120 SPH2 coupled with a Nanometrics Centaur CTR4-6S, 6-ch 24-bit data logger. The data cover the period from January 2022 until July 2025 with a duty cycle $>90\%$. The median of the ASD distribution on 120 s long spectral segments covering this time span for P2 shown in Figure \ref{fig:overall_spectra}. In the same figure, we also compare the data from Sardinia against the data from a borehole seismometer at -250 m installed in the village of Terziet (The Netherlands) for ET site characterization purposes. The data from Terziet are not necessarily representative of the entire EMR candidate region.
Figure \ref{fig:overall_spectra} highlights the quietness of the Sardinia site, with the seismic noise levels within 10$\%$ of Peterson's New Low Noise Model (NLNM) \cite{peterson1993observations} between 4 Hz and 6.5 Hz. This is an outstanding property in the framework of experiments in low noise environments. In fact, the NLNM is a global statistical property, which represents the lower limit of seismic noise on Earth obtained from the 10th percentile of the distribution of seismic noise spectra from seismic stations all around the globe. As a consequence, Figure \ref{fig:overall_spectra}  testifies that Sardinia can reach seismic noise levels which are among the quietest on Earth, suitable to host fundamental physics experiment in low environmental noise conditions \cite{digiovannietal2021, digiovanni2023}. This is confirmed also by Figure \ref{fig:ranking}, which provides the ranking of P2 on the global scale and compares it against other borehole stations that compose the Incorporated Research Institutions for Seismology virtual network FDSN+ALL. 

To evaluate the seasonality of seismic noise levels \cite{digiovanni2023} from P2 to be passed as input to Equation \ref{eq:NN}, we calculate separately the global spectra for each month between 2022 and 2025. Overall, we found that the quietest month is July, whereas the noisiest is December (Figure \ref{fig:season_spectra}) \cite{digiovanni2023}, which also exhibits the larger variability of seismic noise. In fact, not only its median noise level is higher than the remaining months, but its 95th percentile exceeds the global equivalent of Figure \ref{fig:overall_spectra} and of Figure 3 of \cite{DiGiovanni2025}. As a consequence, we consider this as an outlier and December as representative of the worst possible scenario in terms on variability of seismic noise due to environmental effects. On the other hand, we consider the 5th percentile in July as the overall best case scenario. These are used to compare the impact of seasonality over the performance of ET (Figure \ref{fig:sens_spectra}).

\section{Impact on ET design sensitivity and Signal to Noise Ratio}
\label{sec:snr}
 \begin{figure}[h]
     \centering
     \includegraphics[width=0.8\linewidth]{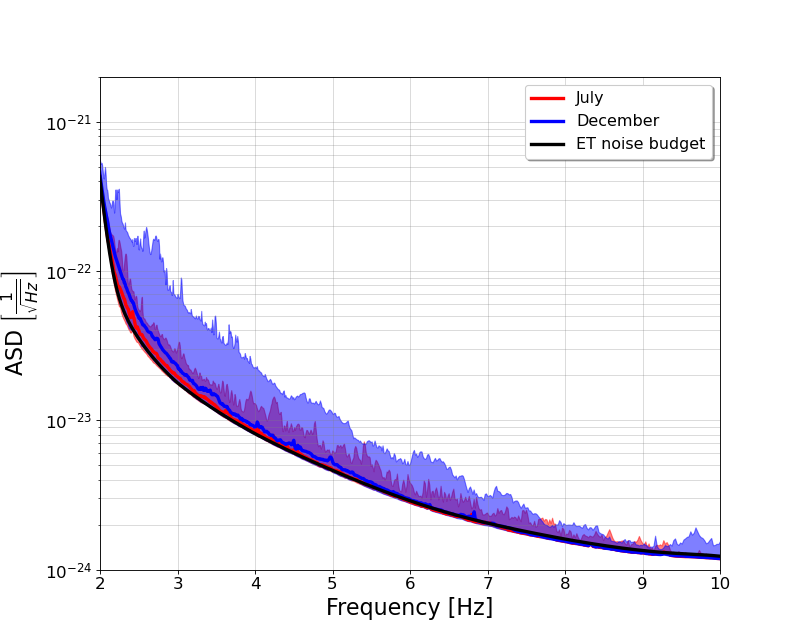}
     \caption{The impact of seasonality on the design sensitivity of ET (black curve). The seasonal changes are represented by the red (July) and black (December) curves. The shaded regions represent the area included between the 5th and 95th percentiles.}
     \label{fig:sens_spectra}
\end{figure}
\begin{figure}[h]
     \centering
     \includegraphics[width=0.8\linewidth]{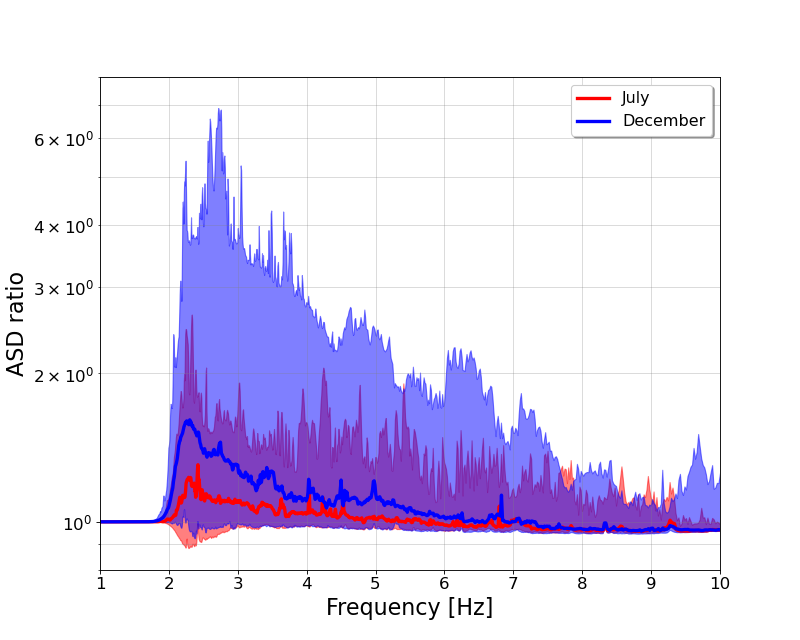}
     \caption{Ratio between the modified sensitivity curves for July (red) and December (blue) and the design sensitivity curve. The shaded regions represent the area included between the 5th and 95th percentiles. Considering only the median noise levels, up to 6 Hz the sensitivity is degraded by, at worst, a factor 1.6, whereas beyond 6 Hz Sardinia performs better than the design case. The larger departure from the design case happens when considering the worst possible scenario: the 95th percentile in December.}
     \label{fig:sens_ratio}
\end{figure}
Starting our analysis from the median seismic noise levels, we note that our findings are consistent with \cite{DiGiovanni2025}, the general quietness of the Sardinia site \cite{digiovannietal2021, digiovanni2023, DiGiovanni2025} translates into a minimal impact on the design sensitivity of ET (Figure \ref{fig:sens_spectra}). Figure \ref{fig:sens_ratio} also shows the ratio between the modified sensitivity curves for July and December and the ET design sensitivity. Considering the medians, up to 6 Hz the design sensitivity is degraded by at worst a factor 1.6 at 2.3 Hz in December. Beyond 6 Hz, Sardinia performs slightly better than the requirements for ET, regardless of the season. The sensitivity curves from the median noise levels in July and December are also consistent with Figure 5 of \cite{DiGiovanni2025} calculated between 2022 and 2023, with only December being slightly noisier.

Shifting our attention to the effect of the variability of the seismic noise levels over the ET sensitivity, shown as colored bands in Figures \ref{fig:sens_spectra} and \ref{fig:sens_ratio}, we note a larger impact. In the best possible case (5th percentile in July) the sensitivity can be up to a factor 1.12, at 2.3 Hz, better than expected (Figure \ref{fig:sens_ratio}). On the contrary, in the worst possible case (95th percentile in December) the worsening of the design sensitivity peaks at a factor 6.9 at 3.6 Hz. On top of this discussion, we remember that in Equation \ref{eq:NN} we are not considering any NN mitigation factor. As a consequence, identifying the frequency regions of higher departure from the expected design sensitivity can provide relevant information for the design of optimized noise mitigation strategies and sensors as well as for how to protect the noise environment during seasonal changes.

\begin{figure}
     \centering
     \includegraphics[width=0.8\linewidth]{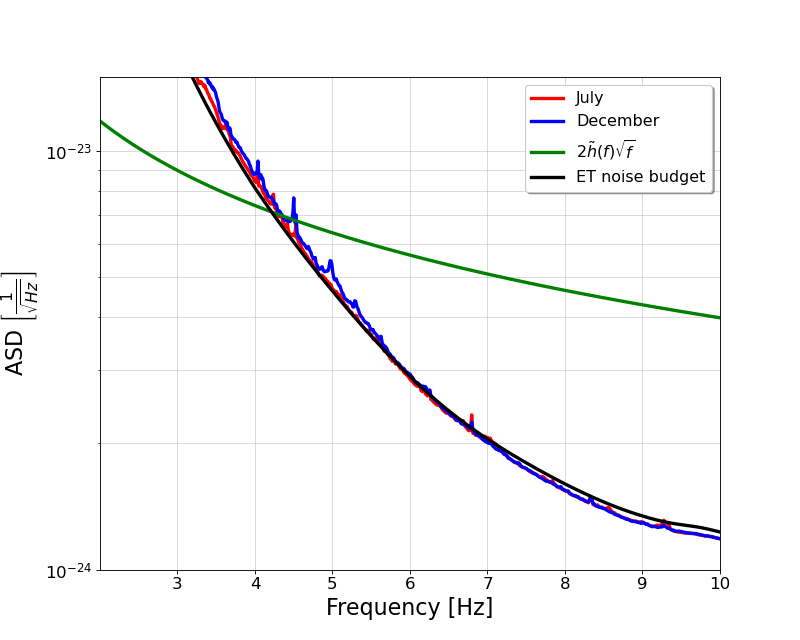}
     \caption{Characteristic strain $2\tilde h(f)\sqrt{f}$ of a sample BNS from \cite{cobacatalog} at 16 Gpc with $M_\mathrm{tot} = 3.47 \, M_{\bigodot}$ overlapped with the median sensitivity curves used in this work.}
     \label{fig:sens_spectra_signal}
\end{figure}

\begin{figure}[h]
     \centering
     \begin{subfigure}[]{
         \centering
         \includegraphics[width=70mm]{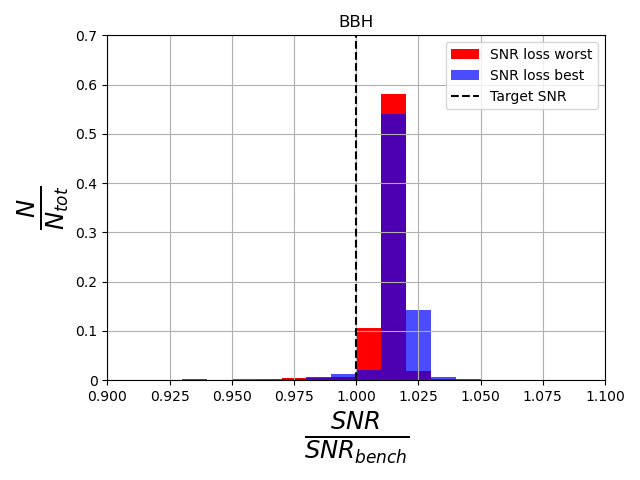}}
         \label{fig:snra1}
     \end{subfigure}
     \begin{subfigure}[]{
         \centering
         \includegraphics[width=70mm]{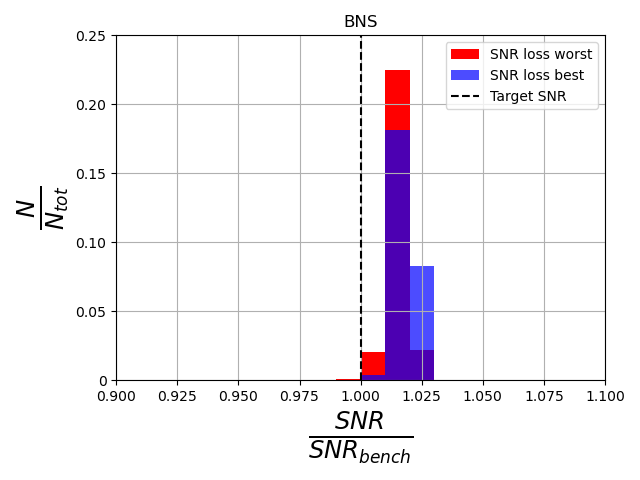}}
         \label{fig:snra2}
     \end{subfigure}
        \caption{SNR loss distributions for IMBH (a) and BNS (b) cases obtained using the 50th percentile of the ground motion PPSDs from July and December. The target SNR is shown as a vertical black dashed line.}
        \label{fig:snr_dist_imbh}
\end{figure}

Figure \ref{fig:snra1} shows the distributions of the ratio between the measured SNR and the SNR calculated for the design sensitivity case ($\mathrm{\frac{SNR}{SNR_{DESIGN}}}$) for the IMBH case. These distributions show any loss/gain of SNR with respect to the design sensitivity case. The events that never reach the threshold $SNR = 12$ in any case, are removed from the analysis. The overall results, including all classes of astrophysical objects considered in this work, are also summarized in Table \ref{tab:BNS_summary}. The consistency with the design case is emphasized by an average gain in SNR of less than 2\% in both July and December. This slight increase, in spite of the degradation of the sensitivity below $\simeq 5.6$ Hz, is explained by Figure \ref{fig:sens_spectra_signal}. Here we show the characteristic strain $2\tilde h(f)\sqrt{f}$ of one of the analyzed BNS signals, placed at 16 Gpc with $M_\mathrm{tot} = 3.47 \, M_{\bigodot}$, compared against the sensitivity curves used in this work. This figure also shows that the larger contribution to the SNR comes from the portion of the bandwidth in which the modified sensitivity is actually better than design, thus explaining why we get a small gain in the SNR of the signals with respect to the design case. On the whole, these results are consistent with Figures 8 and 9 of \cite{DiGiovanni2025}, which also considered lower noise levels.
Moreover, as expected, the analysis for the BNS case (Figure \ref{fig:snra2}) delivers a performance consistent with the IMBH case.
\begin{figure}[h]
     \centering
     \begin{subfigure}[]{
         \centering
         \includegraphics[width=70mm]{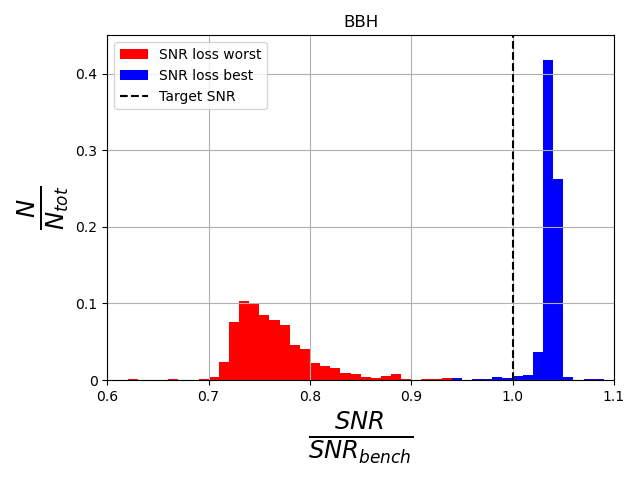}}
         \label{fig:snrb1}
     \end{subfigure}
     \begin{subfigure}[]{
         \centering
         \includegraphics[width=70mm]{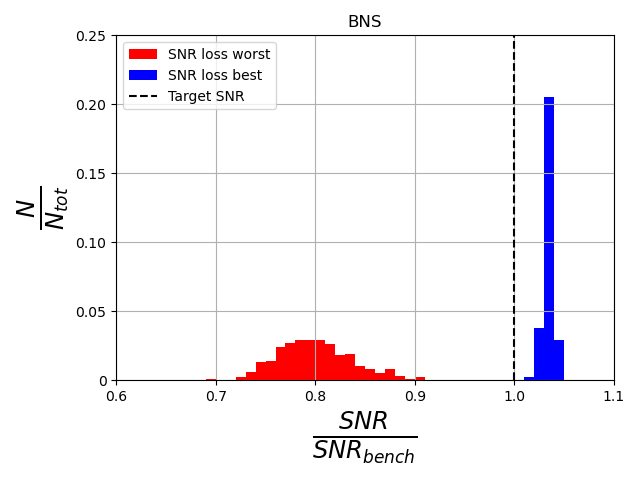}}
         \label{fig:snrb2}
     \end{subfigure}
        \caption{SNR loss distributions for IMBH (a) and BNS (b) cases obtained using the best and worst possible case scenarios from July and December. The target SNR is shown as a vertical black dashed line.}
        \label{fig:snr_dist_bbh_worstbest}
\end{figure}

On the other hand, if we consider the maximum range of variability of the noise levels represented by the best and worst case scenarios, the picture is somewhat altered. If we consider the best case selected for this work (5th percentile of seismic noise levels in July), we note another increase in the SNR of the signals with a mean value of 3.6\% (Figures \ref{fig:snrb1} and \ref{fig:snrb2}). On the other hand, as expected, the most significant loss happens in the worst case (95th percentile of seismic noise levels in December). Here, we see, on average, a loss of 23\% of the SNR of the signals (Figures \ref{fig:snrb1} and \ref{fig:snrb2}). The results for the BNS and IMBH case are totally consistent, with the only difference being in the number of events reaching the SNR = 12 threshold, due to the signals being intrinsically fainter.

\begin{table*}
\centering
\caption{Summary of the SNR performance for the Sardinia and EMR candidate sites. 
The figures reported in this table are the average of the total IMBH and BNS cases. All events which never reach SNR = 12 in any case, are discarded.
The separate tables for the two classes of events are reported in the appendix. 
The negative figures under the events with SNR $<12$ highlight an improvement in the fraction of recovered events with respect to the design case.}
\label{tab:BNS_summary}

\begin{tabular}{lccccc}
\toprule
 & \multicolumn{2}{c}{SNR/SNR$_{\mathrm{DESIGN}}$} 
 & \multicolumn{2}{c}{EVENTS WITH SNR $< 12$}  \\
\cmidrule(lr){2-3} \cmidrule(lr){4-5}
Case & IMBH & BNS & IMBH & BNS & \\
\midrule
Design & -   & - & 24 &  19 \\
July (median) & $+1.6\%$   & $+1.8\%$& 0 &0\\
July (5th perc.) & $+3.6\%$ & $+3.5\%$&0 &0\\
December (median) & $+1.1\%$ & $+1.5\%$&5 &5\\
December (95th perc.) & $-24\%$ & $-20\%$&312 &165\\
\bottomrule
\end{tabular}

\end{table*}

\section{Conclusions and future plans}
\label{sec:conc}
In this work, we quantify the impact of seasonal variability in seismic noise on the LF performance of the Einstein Telescope at the Sardinia candidate site. Using borehole seismic data acquired between 2022 and 2025, we identified representative best- and worst-case conditions, corresponding to July and December, and propagated these into modified sensitivity curves through an estimate of the Newtonian noise contribution in the 2–10 Hz band.
Our analysis confirms that the Sardinia site, on average, is characterized by intrinsically low seismic noise, which translates into a limited seasonal modulation of the detector sensitivity. When considering median noise levels, deviations from the design sensitivity remain modest and are confined to frequencies below approximately 6 Hz, with a maximum degradation of order 1.6 at 2.3 Hz. At higher frequencies, the site meets or slightly outperforms the design target. These estimates are conservative, as no Newtonian noise mitigation factor has been included.
The effect of these variations on the detectability of compact binary coalescence signals has been assessed through simulations of binary neutron star and intermediate-mass black hole systems within the ET triangular configuration, including realistic detector geometry and time-dependent antenna response. The resulting impact on the signal-to-noise ratio is found to be minimal: average deviations from the design case remain at the level of a few percent, typically below 2\% when considering median seasonal conditions. Correspondingly, the fraction of events falling below a detection threshold is largely unaffected, indicating that seasonal seismic fluctuations do not significantly impair early inspiral detectability.
Only in extreme conditions, represented by the upper tail of the December noise distribution, does a more noticeable degradation emerge, with average SNR losses at the level of 20–25\%. However, such scenarios are associated with rare fluctuations and are expected to be mitigated, at least in part, by future noise cancellation strategies.
Overall, these results demonstrate that the Sardinia candidate site is robust against seasonal environmental variations and capable of supporting the low-frequency sensitivity goals of the Einstein Telescope. This stability is particularly relevant for science cases that rely on early-time signal accumulation, such as advance warning for binary neutron star mergers and detailed observations of intermediate-mass black hole systems.
Future work will extend this analysis by incorporating surface wave contribution in Equation \ref{eq:NN}, exploring additional environmental noise sources and investigating of the 2L configuration to provide a robust comparison of the performance of this layout with respect to the classical ET triangle. 

\begin{appendices}

\label{appendix:appendix}
\renewcommand\thefigure{A.\arabic{figure}}    
\renewcommand\thesection{A.\arabic{section}}
\renewcommand\theequation{A.\arabic{equation}}
\setcounter{figure}{0} 
\setcounter{section}{0}
\setcounter{equation}{0}
\section{Derivation of Equation 1}
Since previous works referencing Equation \ref{eq:NN} \cite{alloccaetal2021, harms2022, janssens2022, janssens2024, DiGiovanni2025} did not provide a clear derivation of it, here we  deliver its demonstration for the sake of clarity. First, we consider the strain of a GW interferometer as the fractional change in length of the arms of the detector with respect to its baseline \textit{L}
\begin{equation}\label{eq:A1}
    h = \frac{\Delta L}{L}.
\end{equation}
$\Delta L$ is the relative difference in the length of the two arms. Since the light beam travels back and forth between the test masses, the physical displacement $\Delta L$ of one mirror changes the total optical path length by $2\Delta L$. At this point, for the estimation of the NN contribution to the GW strain, we consider only the $\Delta L$ due to the ground motion field $\xi (\vec r, t)$, which we assume to be stationary, homogeneous and not correlated between the test masses. As such, the power spectral density (PSD) of GW strain in Equation \ref{eq:A1} can be written as
\begin{equation}\label{eq:sdelta}
    S_{h} = S_{\frac{\Delta L}{L}} = \frac{4}{L^2} S_{\Delta L},
\end{equation}
Where $S_{\Delta L}$ represents the PSD of the displacement caused by ground motion only in a single detector.
It is worth pointing out that the assumption of uncorrelated seismic ground motion between the test masses of a single interferometer may not always hold, even if 10 km apart. Indeed, further site characterization studies are currently investigating this aspect and the model will be updated when more detailed information will become available.

Now, we take account of the acceleration produced on a test mass at the center of a spherical cavern in an infinite homogeneous medium by ground motion \cite{harms}:
\begin{equation}\label{eq:deltaacc}
    {\tilde a}(\vec r,\omega) = \frac{4\pi}{3}G\rho_0\left[2{\xi_P}(\vec r,\omega)-{\xi_S}(\vec r,\omega)\right],
\end{equation}
where $\xi_{P,S}(\vec r,\omega)$ is the Fourier amplitude of the seismic displacement vector for the compressional (P) and shear (S) waves, respectively, and $\omega$ is the angular frequency. The derivation of Equation \ref{eq:deltaacc} can be found in Section 3.3 of \cite{harms}. We convert Equation \ref{eq:deltaacc} into displacement and equate its square to the power spectrum at the second term of \ref{eq:sdelta} and get:
\begin{equation}\label{eq:deltacc2}
    S_{\Delta L} = \left[\frac{\tilde a(\vec r,\omega)}{\omega^2}\right]^2 = \left \{ \frac{4\pi}{3}\frac{G\rho_0}{\omega ^2}\left[2{\xi_P}(\vec r,\omega)-{\xi_S}(\vec r,\omega)\right]\right \}^2 = \left(\frac{4\pi}{3}\frac{G\rho_0}{\omega ^2}\right)^2S_{2{\xi_P}-{\xi_S}}.
\end{equation}
Moreover, we note that the PSD of the measured ground motion $S(\xi, \omega)$ can be expressed as the superposition of P and S waves contributions by defining a mixing parameter $p$: $S(\xi_P, \omega) = pS(\xi, \omega)$ and $S(\xi_S, \omega) = (1-p)S(\xi, \omega)$. As a consequence, assuming the lack of coherence which cancels the cross term $4\cdot\mathrm{Re}S_{\xi_P\xi_S}$,
\begin{equation}\label{eq:delta_acc3}
    {S_{2{\xi_P}-{\xi_S}}} = {4S(\xi_P, \omega) + S(\xi_S, \omega)} = {(1+3p)S(\xi, \omega)}
\end{equation}
and Equation \ref{eq:deltacc2} becomes
\begin{equation}
    S_{\Delta L}=\left(\frac{4\pi}{3}\frac{G\rho_0}{\omega ^2}\right)^2{(1+3p)S(\xi, \omega)}.
\end{equation}
Recalling Equation \ref{eq:sdelta}, we get
\begin{equation}
S_{h} = \frac{4}{L^2} S_{\Delta L} = \frac{4}{L^2}\left(\frac{4\pi}{3}\frac{G\rho_0}{\omega ^2}\right)^2{(1+3p)S(\xi, \omega)}
\end{equation}
Assuming that the energy partition spreads uniformly across the degrees of freedom of body waves (1 for P waves, 2 for S waves) we set p = 1/3. This assumption is made because the composition of the seismic fields is not known yet for the sites of interest, and so an ad hoc assumption was made that displacement is equally distributed among the three polarizations of seismic waves \cite{harms, harms2022}. The actual value  might be different for each site and depend on the types of seismic sources, their distance, and the amount of scattering that affects a wavefield. Future site characterization activities will solve this debate. If we return to the amplitude domain and define $\tilde x(\omega) = \sqrt{S(\xi, \omega)}$ as the amplitude spectral density of measured ground motion, we obtain
\begin{equation}
    {h}_{\rm{NN}}(f) = \frac{4\pi}{3}G\rho_0\frac{2\sqrt{2}}{L}\frac{1}{(2\pi f)^2}\tilde{x}(f),
\end{equation}
which is Equation \ref{eq:NN}.
\section{Monthly Seismic Spectra}
\begin{figure}[htbp]
     \centering
     \includegraphics[width=0.8\linewidth]{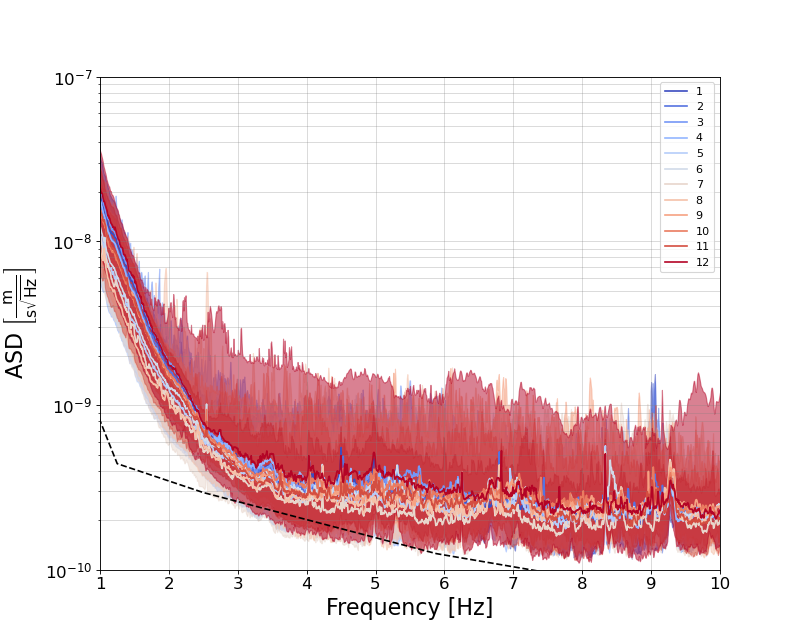}
     \caption{Monthly spectra evaluated for this work to extract the best and worst case scenario.}
     \label{fig:all_months}
\end{figure}

%
In Figure \ref{fig:all_months} we show the entire set of the monthly spectra calculated for the 2022-2025 time span over 120 s long segments considered for the analysis presented in this work. It is apparent that the quietest month is July: the median of the seismic noise levels is the lowest as well as the 5th percentile. On the other hand, December clearly represents the worst possible scenario. Except in the frequency intervals 5 Hz - 6.3 Hz and 7.5 - 8.2 Hz, which is slightly below other months, the median of the seismic noise levels dominates over the other periods of the year. Moreover, its 95th percentile is a clear outlier with respect to other months, especially in the 2 Hz - 5 Hz frequency band, proving a good base for a worst case estimate. These figures are naturally in agreement with the seasonality variation analysis already presented in \cite{digiovanni2023}, which focused on seismic noise of natural origin.

\end{appendices}


\end{document}